\documentclass[amsmath,showpacs,twocolumn,aps,prl,superscriptaddress]{revtex4}
\usepackage{graphicx}
\usepackage{dcolumn}
\usepackage{bm}

\newcommand{\bq}{\begin{equation}}
\newcommand{\ee}{\end{equation}} \newcommand{\fr}[2]{\frac{#1}{#2}}

\begin{document}

\title{Mesoscopic Spin-Hall Effect in 2D electron systems with smooth
boundaries}
\author{P. G. Silvestrov}
 \affiliation{Theoretische
Physik III, Ruhr-Universit{\"a}t Bochum, 44780 Bochum, Germany}

\author{V. A. Zyuzin$^*$}
\affiliation{Department of Physics, University of Utah, Salt Lake
City, Utah 84112, USA}
\author{E. G. Mishchenko}
\affiliation{Department of Physics, University of Utah, Salt Lake
City, Utah 84112, USA}

\begin{abstract}

Spin-Hall effect in ballistic 2D electron gas with Rashba-type
spin-orbit coupling and smooth edge confinement is studied. We
predict that the interplay of semiclassical electron motion and
quantum dynamics of spins leads to several distinct features in spin
density along the edge that originate from accumulation of turning
points from many classical trajectories. Strong peak is found near a
point of the vanishing of electron Fermi velocity in the lower
spin-split subband. It is followed by a strip of negative spin
density that extends until the crossing of the local Fermi energy
with the degeneracy point where the two spin subbands intersect.
Beyond this crossing  there is a  wide region of a smooth positive
spin density. The total amount of spin accumulated in each of these
features exceeds greatly the net spin across the entire edge. The
features become more pronounced for shallower boundary potentials,
controlled by gating in typical experimental setups.

\end{abstract}

\pacs{ 73.23.-b, 72.25.-b}

\maketitle

{\it Introduction}. Spin-Hall effect \cite{DP}, manifested in the
boundary spin polarization when electric current flows through a
system with significant spin-orbit interaction, was recently
observed in both 3D \cite{exp1,Kato,exp3} and 2D systems
\cite{exp2}. Two mechanisms that lead to the effect are typically
distinguished. The {\it extrinsic} mechanism is dominant in 3D
semiconductors and originates from scattering off
impurities~\cite{H,Z,ERH,DS}. {\it Intrinsic}
mechanism~\cite{MNZ,Sinova} of the band-structure induced spin
precession can be  realized in ballistic (disorder-free) 2D systems.

The intrinsic spin-Hall mechanism
is of particular appeal. However, in two-dimensional electron
systems with spin-orbit coupling linear in momentum (typical
for $n$-doped heterostructures) any scattering  that
leads to a stationary electric current via deceleration of
electrons by impurities, phonons, etc., will negate the precession
due to external electric field and result in
the exact cancellation \cite{IBM,MSH,Kh,RS,D} of the bulk
spin-current in a dc case \cite{comment}.

One possible way to avoid this cancellation is to move into the ac
domain with frequencies exceeding the inverse spin relaxation time
\cite{MSH}. Another possibility is to use dc fields but  make a
system sufficiently small and clean (ballistic) so that the electron
mean free time exceeds the time of flight across the systems. The
corresponding scenario became known as the \emph{mesoscopic
spin-Hall effect} \cite{nik}. While initial theories of spin-Hall
effect in infinite systems had addressed such auxiliary quantity as
spin current (for a review see Refs.~\cite{reviews,ERH1}), finite
geometry calls for calculations of spin polarization,  a directly
measurable quantity~\cite{nonlocal}.

It is important to emphasize that the edge spin polarization in
ballistic systems appears not as a result of electric field-driven
acceleration of electrons and associated with it precession of
spins.  Instead, it originates from their precession in the course
of electron motion in the boundary potential that provides lateral
confinement. When populations of left- and right-moving states are
different (due to the applied bias) the net effect of this
precession results in a non-zero  spin polarization. Such edge
polarization was considered, mostly by  numerical methods, in
several earlier publications~\cite{NWS,nik,nik3,usaj,SG}.

\begin{figure}[h]
\includegraphics[width=7.cm]{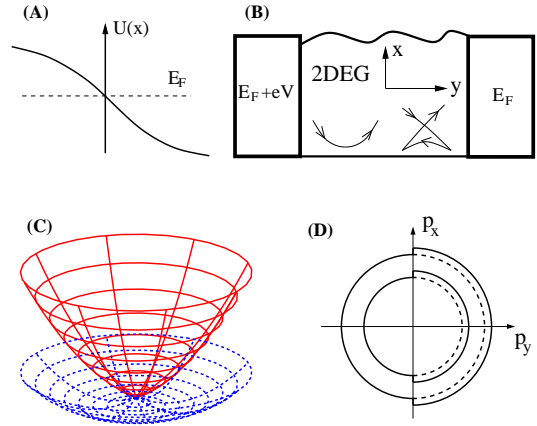}
\caption{A)
Profile of a boundary potential $U(x)$. B) Geometry of the system:
two-dimensional electron gas ($x>0$) is attached to two ideal
reflectionless metallic leads filled by equilibrium electrons up
to different chemical potentials. Two types of boundary scattering
are shown schematically. C)
Spin-orbit-split subbands structure. D) Difference in population
of left- and right-moving electrons due to the applied bias $eV$.
} \label{fig1}
\end{figure}

The total amount of accumulated spin near an edge of a 2DEG is {\it
independent} of the boundary potential, as found in our recent
study~\cite{ZSM}, yet only second order in the spin-orbit coupling
constant. In the present paper we show how to overcome this
limitation and achieve a much larger local spin-Hall polarization
over extended strips along the edge of 2DEG provided that {\it
smooth} boundaries are utilized. Increasing the width of the edge by
making the boundary potential progressively smoother, one can
increase the amount of spin accumulated within each strip. The sign
of the polarization in these strips will alternate so that the net
spin accumulation will be in agreement with Ref.~\cite{ZSM}.

{\it Spin dynamics in smooth potentials.} Consider ballistic
 two-dimensional electron gas (2DEG) attached
to two ideal perfectly conducting leads which are kept under a
voltage bias $V$, see Fig.~\ref{fig1}. Hamiltonian of the system
features spin-orbit coupling of the usual ``Rashba''
type~\cite{BR},
 \begin{equation}
\label{Ham}
 \hat{H}=\frac{\hat p_x^2+\hat p_y^2}{2m}+\frac{2\lambda}{\hbar}(\hat s_x
\hat p_y-\hat s_y \hat p_x)+\frac{m\lambda^2}{2}+U(x).
 \end{equation}
We assume that the spin-orbit coupling strength $\lambda$ is the
same inside 2DEG and in the leads (or, equivalently, that
switching-on of $\lambda$ happens adiabatically as electrons travel
from the leads towards 2DEG). In this case the applied bias
transforms into the difference of Fermi energies for the left and
right moving electrons far from the edges of 2DEG,
$E_{F_R}-E_{F_L}=eV$.

The density of the out-of-plane spin $s_z$ can be related to the
dynamics of the in-plane spin via simple identities,
 \bq\label{commutator1}
\fr{d}{dt}\psi^\dagger \hat{s}_y\psi
=\fr{i}{\hbar}\psi^\dagger[\hat{H},\hat{s}_y]\psi -\nabla
\vec{J}^y  \ , \ [\hat{H},\hat{s}_y]=2i\lambda \hat{s}_z p_y.
 \ee
Here the operator of spin current $ \vec{J}^y = \fr{i\hbar}{2m}
[(\nabla \psi)^\dagger \hat{s}_y\psi -\psi^\dagger
\nabla\hat{s}_y\psi] +\fr{2\lambda}{\hbar}\psi^\dagger
\vec{a}\hat{s}_y \psi$, and $\vec{a}=(-\hat{s}_y,\hat{s}_x)$.
Since the momentum along the edge $p_y$ is conserved, and we are
interested in the stationary spin-density, we find
 \bq\label{s_accumulated}
{\cal{S}}(x)=\sum_i\int_{-\infty}^x \psi^\dagger_i (x') s_z\psi_i
(x') dx' = \sum_i\fr{-\hbar}{2\lambda p_y}J^y_{xi}(x),
 \ee
Here the sum is extended over all occupied states. The exact value
of the out-of-plane spin is thus determined by the local current
$J^y_x(x)$. Next, we notice that in a smooth potential $U(x)$ the
subband index is almost conserved. The electron's spin remains
{\it largely in-plane}, while being also perpendicular to the
momentum, with the local value of $p_x(x)$ determined from the
energy conservation. One can now introduce the {\it
nonequilibrium} (Fig.~1D) electron distribution for the local
value of the boundary potential $U(x)$ and the  Fermi velocity
$v_F(x)=\sqrt{-2U(x)/m}$ and calculate the expectation value of
$J^y_x(x)$. This yields,
 \begin{equation} \label{s_density}
\overline {\cal{S}}(x) =\frac{eV}{2\lambda(2\pi)^2}
\left(\frac{2\lambda}{v_F(x)} -
\ln{\frac{v_F(x)+\lambda}{|v_F(x)-\lambda|}}\right).
 \end{equation}
Knowledge of the local distribution, however, does not allow one
to describe {\it quantum oscillations} due to the interference of
waves with different $p_x$ contributing to the same eigenfunction
$\psi_i$. Thus Eq.~(\ref{s_density}) describes only  a smooth
semiclassical part of spin accumulation, as denoted by the bar,
$\overline {\cal{S}}$. We note that the accumulated spin ${S}(x)$
is a convenient quantity for numerical calculations, since the
integration over $x$ reduces the relative weight of the quantum
oscillations. Differentiating Eq.~(\ref{s_density}) over $x$ we
find that the spin density is proportional to the local value of
the force exerted by the boundary potential,
 \begin{equation}\label{local_spin}
\overline{s_z} (x) = - \frac{\lambda^2 eV}{(2\pi)^2m
(v_F^2(x)-\lambda^2)v_F^3(x)}\frac{dU}{dx}.
 \end{equation}
Note that the net spin polarization across the edge is independent
of the shape of the boundary potential, $ {\cal{S}}(\infty) = -
{\lambda^2 eV}/{12\pi^2 v^3_F}$, and is expressed via the bulk value
of the Fermi velocity $v_F(\infty)$. Here by $x=\infty$ we assume a
point deep inside the 2DEG but yet far from its opposite edge. The
latter has spin accumulation of the same absolute value and opposite
sign.

\begin{figure}[h]
\includegraphics[width=8.0cm]{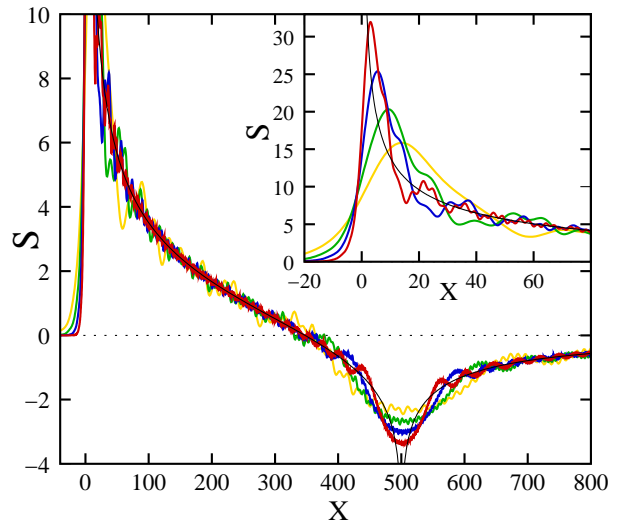}
\caption{Integrated spin density  ${\cal S}(x)=\int_{-\infty}^x s_z
dx$ for the potential $U(x)=-\alpha m^2\lambda^3 x/\hbar$ in units
of $eV/8\lambda\pi^2$. The curves for $\alpha = 8, 4, 2, 1 \times
10^{-3}$ are shown in yellow, green, blue and red respectively. The
horizontal coordinate is scaled differently for different curves, as
$x$ is measured in units of $10^3\alpha \times \hbar/m\lambda$.
Narrow black lines stand for the semiclassical prediction,
Eq.~(\ref{s_density}). The logarithimc behavior, $\sim \log \alpha$,
of the dip at at $U(x)=-m\lambda^2/2$ ($x=500$) is clearly seen.
Inset magnifies the region near the edge of 2DEG ($x\approx 0$).}
\label{fig2}
\end{figure}

{\it Numerical solution.} To illustrate the predicted dependence
(\ref{s_density}-\ref{local_spin}) we performed exact numerical
calculation of the spin density accumulated near the boundary
approximated by a linear potential $U(x)=-Fx$ with the constant
force $F$. The smoothness of the boundary implies that $F\ll
m^2\lambda^3/\hbar$. Fig.~\ref{fig2} demonstrates an excellent
agreement between the predicted integrated spin density
Eq.~(\ref{s_density}) and exact numerical simulations for
different values of the slope of the boundary potential.

According to Eq.~(\ref{s_density}) we find two regions of
different smooth spin behavior. First, within the strip where
$0<v_F(x)<\lambda$ spin density is negative (which is seen in a
downward slope of the integrated density ${\cal{S}}(x)$ in
Fig.~\ref{fig2}). Farther away, $s_z(x)$ changes sign for
$v_F(x)>\lambda$, where both $s_z(x)$ and ${\cal{S}}(x)$  decrease
gradually with increasing~$x$.

The most interesting is the behavior of spin  at the borders of
these regions, $v_F=0$ and $v_F=\lambda$. At $v_F(x)=\lambda$ the
accumulated spin $\overline {\cal{S}}(x)$ Eq.~(\ref{s_density})
diverges logarithmically. This singularity originates from the
accumulation of classical turning points taking place when the
conical crossing point in the spectrum of the Hamiltonian
(\ref{Ham}), see Fig.~1C, passes through the Fermi energy. This
singularity is regularized as ${\cal{S}}\sim \log F$, according to
Fig.~\ref{fig2}.

Yet more peculiar is the behavior of both $s_z(x)$ and
${\cal{S}}(x)$ at the edge of 2DEG, near the point where $v_F(x)=0$.
The smooth part of the accumulated spin, Eq.~(\ref{s_density}), has
an infinite jump here (from Eq.~(\ref{s_density}) it follows that
$\overline {\cal{S}}(+0)=\infty$, while of course $\overline
{\cal{S}}(-0)=0$). Development of such jump with decreasing slope of
the potential is seen in the inset in Fig.~\ref{fig2}. The jump in
${\cal{S}}(x)$ corresponds to the formation of a narrow strip with
extremely large values of spin $s_z>0$ along the border. This
behavior will now be analyzed in more detail.

\begin{figure}
\includegraphics[width=8.0cm]{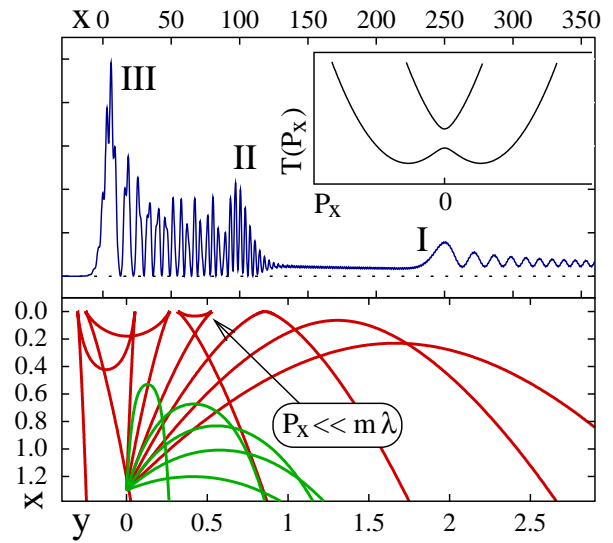}
\caption{ {\it Bottom:} Family of classical trajectories at
$E=E_F$ for different values of $p_y$, and $m=\lambda=F$.
Trajectories for both lower (red) and upper (green) spin-orbit
split subbands are shown. An example of a trajectory contributing
to the peak in spin density at $x\approx 0$, Eq.~(\ref{stop}), is
shown for $"p_x\ll m\lambda"$ (see the text). {\it Top:} The
electron density $\rho =\psi_1^\dagger \psi_1 + \psi_2^\dagger
\psi_2$ for  given longitudinal momentum and energy. Here
$\psi_{1,2}$ are two eigenfunctions of the Hamiltonian
Eq.~(\ref{Ham}) with $p_y=0.2m\lambda, E_F=0,
F=0.003m^2\lambda^3/\hbar$, $\rho$ in arbitrary units, $x$ in
units of $\hbar/m\lambda$. Three classical turning points can be
seen. The interference of incoming and reflected waves in the
upper subband causes smooth oscillations to the right of the inner
turning point I ($x>m\lambda^2/2F$). At the other turning points,
II and III, the two kinds of oscillations are seen. Slow
oscillations are caused by the interference of the incoming wave
and the wave reflected at the turning point. Fast oscillations
(wavelength $\sim\hbar/m\lambda$) are due to the interference of
distant (in time) segments of the same trajectory. {\it Inset:}
Kinetic energy (arbitrary units) $T_\pm(p_x)=(p\pm m\lambda)^2/2m$
for fixed $p_y=0.2 m\lambda$.} \label{fig3}
\end{figure}

{\it Semiclassical analysis}. Classical dynamics of electrons with
Rashba spin-orbit interaction is described by the Hamilton function
 \cite{SM}
 \bq\label{Heff}
{\cal H}_{\pm}=\fr{(p \pm m\lambda)^2}{2m} -Fx,
 \ee
 for the two spin-split subbands. The boundary potential
 in Eq.~(\ref{Heff}) is again approximated by the linear function.
The family of classical trajectories generated by the Hamilton
function Eq.~(\ref{Heff}), shown in Fig.~\ref{fig3}, demonstrate a
number of unusual features.

As seen from Fig.~\ref{fig3}, those electrons from the lower
subband ($-$) that have $|p_y|<m\lambda$, pass {\it three} turning
points in the course of their motion in the $x$ direction,
corresponding to three solutions of the equation $\partial {\cal
H}_{-}/\partial p_x=0$. Two of these turning points (those with
$p=m\lambda$) correspond to simultaneous vanishing of {\it both}
velocity components, $\vec{v}(x)=0$, the behavior generically
impossible in a 2DEG with the parabolic spectrum, ${\cal
H}=p^2/2m$. The fundamental difference in the classical dynamics
of spin-orbit-split subbands lies with the fact that all electron
states in the lower subband with the {\it same energy} but {\it
different momenta} stop at these two turning points {\it at the
same point} $x_0$, $U(x_0)=E_F$, provided that $|p_y| \le
m\lambda$. Consequences of this fact for the anomalous behavior of
the ballistic conductance have been discussed in Ref.~\cite{SM}.

Both singularities in the spin density (\ref{local_spin}),
$v_F(x)=0$ and $v_F(x)=\lambda$, originate from the accumulation
of classical turning points from many trajectories.  Let us now
demonstrate how the stronger of the two singularities, $v_F(x)=0$,
is regularized when the spin density is calculated from the
solutions of the Schr\"{o}dinger equation with the Hamiltonian
(\ref{Ham}). We first perform  decomposition of the wavefunction
$\psi(x,y)$ into a product of the fast exponent $e^{i\vec{p}\cdot
\vec{r}}$ and a slow spinor function. Components of the latter
satisfy the Schr\"odinger equation for a particle in the
homogeneous field $F$ with the effective mass $m_{\rm eff}=
(m\lambda/p_x)^2 m$. Corresponding solutions are the well known
Airy functions~\cite{LandaFshitz}. At $E=E_F=0$ we find in the
vicinity of the turning point $x=0$,
 \begin{equation}\label{Airy}
\psi(x,y)=m\left(\frac{\lambda^4}{2Fp_x^4} \right)^{1/6}
\left(\begin{array}{c} e^{-i\alpha} \\
i e^{i\alpha}
\end{array} \right)\text{Ai}(\xi)~e^{i\vec{p}\cdot
\vec{r}},
 \end{equation}
where $\xi=-x~m(2F\lambda^2/p_x^2)^{1/3}$,
$p_x^2+p_y^2=m^2\lambda^2$, and $\tan
2\alpha=\sqrt{(m\lambda-p_x)/(m\lambda+p_x)}$.

As seen from Fig.~\ref{fig3}, electron trajectories always bounce
twice at the turning point $x=0$. This leads to the interference
of the two solutions, Eq.~(\ref{Airy}), with the opposite signs of
$p_x$, see Fig.~\ref{fig3} top. Ignoring these microscopic
oscillations that occur on the scale $\sim\hbar/m\lambda$, we can
write the expectation value of the $z$-component of electron spin
for $x\ll m\lambda^2/F$ as follows,
 \begin{equation}
\label{szAiry} s_z =\frac{3meV}{4\pi\hbar}
\frac{\partial}{\partial \widetilde x} \int\limits_0^{1} dz~
\text{Ai}^2(- \widetilde x/z),
 \end{equation}
where $\widetilde x=x(2Fm/\hbar^2)^{1/3}$. In the asymptotic
region $x\gg (2Fm/\hbar^2)^{-1/3}$ one can average over the
oscillations of the Airy function. This allows us to recover the
singular behavior of the smooth spin density (\ref{local_spin}):
$\langle s_z\rangle\sim x^{-3/2}$.

The most interesting is the behavior of  spin density close to the
turning point, for $|x|\ll (2Fm/\hbar^2)^{1/3}$. The integral in
Eq.~(\ref{szAiry})  features a logarithmic singularity at $x= 0$.
This enhancement of spin-density is due to the electrons with
$p_x\ll m\lambda$, whose wavefunctions, given by Eq.~(\ref{Airy}),
oscillate rapidly and add {\it coherently} only at the point
$x=0$. With the logarithmic accuracy the height of the peak of
spin density is
\begin{equation}
\label{stop}
s_z(0)=\frac{meV}{10\sqrt{3}\pi^2\hbar}\ln{\left(\frac{m^2\lambda^3}{\hbar
F} \right)}.
\end{equation}
Striking feature of this result is that this maximal value is
virtually {\it independent} of the strength of spin-orbit coupling
or the shape of the boundary potential (up to a weak logarithmic
factor).

\begin{figure}[h]
\includegraphics[width=8.5cm]{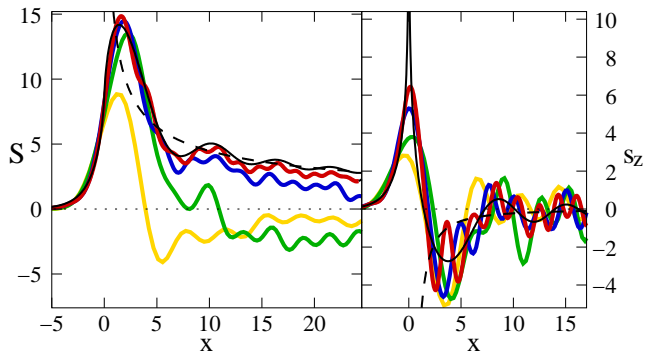}
\caption{Integrated spin  ${\cal S}(x)=\int_{-\infty}^x s_z dx$
(left panel) and spin density $s_z(z)$ (right panel) for $F=-\alpha
m^2\lambda^3 /\hbar$,  in the vicinity of the turning point $x=0$.
Curves for $\alpha = 0.25, 0.05, 0.01, 0.002$ are shown in yellow,
green, blue and red respectively. The distance $x$ is measured in
units of $ (100 \alpha)^{-1/3} \hbar/m\lambda$. Integrated spin
${\cal S}$ and spin density $s_z$ are measured in units of $(100
\alpha)^{1/3}eVm/8\hbar \pi^2$ and $eVm/8\lambda \pi^2$,
respectively. Dashed black lines stand for the semiclassical result
(\ref{s_density}-\ref{local_spin}). The approximation (\ref{szAiry})
valid close to the classical turning point is shown by solid black
lines. Development of the logarithmic maximum of the spin density at
$x=0$ for small values of $\alpha$, Eq.~(\ref{stop}), is clearly
seen in the right panel.} \label{fig4}
\end{figure}

Numerical calculations presented in Fig.~\ref{fig4} illustrate the
emergence of the logarithmic peak when the slope of the boundary
potential $F$ decreases, Eq.~(\ref{stop}).

{\it In Summary,} we have predicted that the nonequilibrium
spin-Hall spin accumulation near a smooth boundary of 2DEG
ballistic conductor with spin-orbit interaction develops a narrow
peak at the edge, with the width $\sim (\hbar^2/m F)^{1/3}$ and
height given by Eq.~(\ref{stop}). It is followed by a slow
non-monotonic decay, see Fig.~\ref{fig2}. This smooth tail of spin
density persists to much larger distances, $\gtrsim m\lambda^2/F$.
The amount of spin accumulated in the peak (found as a maximum of
the function ${\cal S}(x)=\int^x s_z dx$) equals ${\cal S}_{\rm
max} \approx 0.04 eV(m^2/\hbar F)^{1/3}>0.04 eV/\lambda$, where in
the last inequality we utilize the fact that $F<
m^2\lambda^3/\hbar$. We thus conclude that the spin accumulated at
the edge described by a semiclassical boundary potential is
inversely proportional to the strength of spin-orbit interaction
and becomes progressively larger for smoother slopes. This
prediction can be used for experimental observation of spin-Hall
effect in realistic two-dimensional electron systems.

This work was supported by the SFB TR 12  and DOE Grant
No.~DE-FG02-06ER46313.


\begin{references}
\bibitem{comment} This exact cancellation does not occur in 3D-
or 2D hole-systems that feature non-linear spin-orbit couplings.
\end{references}

\end{document}